\documentclass[11pt]{article}

\usepackage[margin=1in]{geometry}
\usepackage{hyperref}
\usepackage{url}
\usepackage{amsmath,amssymb,amsfonts}
\usepackage{graphicx}
\usepackage{booktabs}
\usepackage{multirow}
\usepackage{natbib}
\usepackage{xcolor}
\usepackage{enumitem}

\setcitestyle{authoryear,open={(},close={)}}

\title{Quantifying Ranking Instability Across Evaluation Protocol Axes\\in Gene Regulatory Network Benchmarking}

\author{Ihor Kendiukhov\\
Department of Computer Science\\
University of T\"ubingen, Germany\\
\texttt{kenduhov.ig@gmail.com}}

\date{}

\begin{document}

\maketitle

\begin{abstract}
Benchmark rankings are routinely used to justify scientific claims about method quality in gene regulatory network (GRN) inference, yet the stability of these rankings under plausible evaluation-protocol choices is rarely examined.
We present a systematic diagnostic framework for measuring ranking instability under protocol shift, including decomposition tools that separate base-rate effects from discrimination effects.
Using existing single-cell GRN benchmark outputs across three human tissues and six inference methods, we quantify pairwise reversal rates across four protocol axes: candidate-set restriction (16.3\%, 95\% CI 11.0--23.4\%), tissue context (19.3\%), reference-network choice (32.1\%), and symbol-mapping policy (0.0\%).
A permutation null confirms that observed reversal rates are far below random-order expectations (0.163 versus null mean 0.500), indicating partially stable but non-invariant ranking structure.
Our decomposition reveals that reversals are driven by changes in methods' relative discrimination ability rather than by base-rate inflation---a finding that challenges a common implicit assumption in GRN benchmarking.
We propose concrete reporting practices for stability-aware evaluation and provide a diagnostic toolkit for identifying method pairs at risk of reversal.
\end{abstract}

\section{Introduction}

Mechanistic interpretability for biological foundation models increasingly relies on benchmark ranking to support claims about biological plausibility and downstream utility \citep{cui2024scgpt, theodoris2023transfer, yang2022scbert}. In gene regulatory network (GRN) recovery, methods are compared by their ability to reconstruct known regulatory relationships from single-cell expression data, and the resulting leaderboard positions are used to argue that one model has learned more biologically meaningful structure than another \citep{pratapa2020benchmarking, marbach2012wisdom}. However, the evaluation pipeline involves several protocol choices---which candidate edges to score, which reference network to compare against, how to map gene identifiers, and which tissue context to evaluate in---that are rarely reported or controlled \citep{pratapa2020benchmarking, weber2019essential}.

This is not only a reporting issue. If ranking is unstable under plausible protocol variation, biological decisions can flip: which regulators are prioritized for experimental validation, which mechanistic narrative is emphasized in a publication, and which model is treated as scientifically credible. The field therefore needs explicit ranking-stability diagnostics, rather than only larger metric tables \citep{mangul2019systematic, boulesteix2013plea}.

The ranking-sensitivity problem has a broader context in machine learning evaluation. Benchmark sensitivity to evaluation choices has been documented in NLP \citep{post2018call, marie2021scientific}, computer vision \citep{recht2019imagenet}, and across ML subfields generally \citep{dehghani2021benchmark, demsar2006statistical}. However, systematic quantification of reversal rates and their mechanistic drivers remains underdeveloped in the biological setting. In GRN benchmarking specifically, the positive-label base rate can vary by orders of magnitude across candidate-set restrictions \citep{pratapa2020benchmarking, garcia2019benchmark}, creating a confound between apparent performance and evaluation geometry.

We make three contributions:
\begin{enumerate}[nosep]
\item A diagnostic framework that decomposes ranking shifts into base-rate and discrimination components, clarifying which mechanisms drive observed reversals.
\item A multi-axis empirical quantification of ranking instability across candidate-set, tissue, reference-network, and symbol-mapping shifts in GRN benchmarking.
\item Concrete reporting recommendations and a practical screening tool for identifying method pairs at risk of reversal under protocol variation.
\end{enumerate}

\section{Related Work}

\paragraph{GRN benchmarking methodology.}
Systematic benchmarking of GRN inference methods has been an active area since the DREAM challenges \citep{marbach2012wisdom, prill2010towards}, which established community standards for comparing network inference algorithms on both simulated \citep{schaffter2011genenetweaver, dibaeinia2020sergio} and real expression data. More recently, the BEELINE framework \citep{pratapa2020benchmarking} benchmarked 12 algorithms on curated single-cell datasets, while \citet{garcia2019benchmark} systematically compared resources for estimating transcription factor activities. These studies typically compare methods on fixed evaluation protocols, and the sensitivity of rankings to protocol choices---such as which candidate edges to evaluate, which reference to use, and how to map gene identifiers---has received limited formal attention. Our work addresses this gap by quantifying when and why rankings reverse under protocol variation.

\paragraph{Evaluation sensitivity in machine learning.}
The sensitivity of benchmark conclusions to evaluation choices is a recognized concern across machine learning. \citet{post2018call} demonstrated that BLEU score rankings in machine translation are sensitive to tokenization and normalization choices. \citet{recht2019imagenet} showed that classifiers' relative performance can shift when evaluated on new test sets drawn from the same distribution, while \citet{dehghani2021benchmark} argued that benchmark rankings are often artifacts of incidental experimental choices rather than meaningful capability differences. \citet{marie2021scientific} conducted a meta-evaluation of machine translation research, finding widespread issues with evaluation reproducibility. More broadly, \citet{demsar2006statistical} established best practices for statistical comparison of classifiers across multiple datasets, and \citet{boulesteix2013plea} argued for neutral comparison studies in computational sciences. \citet{weber2019essential} provided guidelines for computational method benchmarking in genomics specifically. Our work applies this perspective systematically to GRN method comparison under protocol shift.

\paragraph{GRN inference methods.}
The GRN inference landscape spans classical expression-based approaches---including tree-based methods like GENIE3 \citep{huynhthu2010inferring} and its scalable variant GRNBoost2 \citep{moerman2019grnboost2}, regulatory-network-specific pipelines like SCENIC \citep{aibar2017scenic}, and information-theoretic approaches---as well as more recent foundation-model-based methods. scGPT \citep{cui2024scgpt} and Geneformer \citep{theodoris2023transfer} have proposed extracting GRN predictions from transformer attention patterns, motivating the need for rigorous evaluation. Pre-trained single-cell language models such as scBERT \citep{yang2022scbert} further expand this space. Our analysis includes both attention-derived and classical expression-based methods, enabling cross-paradigm stability assessment.

\paragraph{Reference networks and gold standards.}
GRN evaluation depends critically on reference networks, which serve as ground truth for computing evaluation metrics. Commonly used resources include DoRothEA \citep{holland2020dorothea}, which integrates multiple evidence sources for transcription factor--target interactions; TRRUST \citep{han2018trrust}, a literature-curated database; OmniPath \citep{turei2021integrated}, which aggregates signaling and regulatory interactions from numerous databases; and STRING \citep{szklarczyk2021string}, which provides protein-protein interaction networks. The Tabula Sapiens atlas \citep{tabulasapiens2022} provides tissue-specific expression context for GRN evaluation. These references encode fundamentally different biological evidence classes, and our reference-shift analysis quantifies how this heterogeneity propagates into ranking instability.

\paragraph{Single-cell data analysis.}
The single-cell transcriptomics revolution \citep{luecken2019current, eraslan2019deep} has generated both the data and the impetus for GRN benchmarking. Standard analysis frameworks such as Scanpy \citep{wolf2018scanpy} provide the preprocessing and analysis infrastructure on which GRN evaluation pipelines are built. Benchmarking efforts in related single-cell tasks---such as trajectory inference \citep{saelens2019comparison} and data integration \citep{luecken2022benchmarking}---have similarly highlighted sensitivity to methodological choices, supporting the broader motivation for ranking-stability analysis.

\section{Diagnostic Framework}
\label{sec:framework}

\subsection{Notation}

Let $\mathcal{M}_m(S, \pi, R)$ denote a scalar evaluation metric (e.g., AUPR) for method $m$ under candidate set $S$, mapping policy $\pi$, and reference network $R$. For two methods $A$ and $B$, define the \emph{margin}:
\begin{equation}
\Delta = \mathcal{M}_A - \mathcal{M}_B.
\end{equation}

For two protocol settings (subscripts 1 and 2), let $\Delta_1$ and $\Delta_2$ denote the respective margins, and define the \emph{margin shift} $\delta\Delta = \Delta_2 - \Delta_1$. A \emph{ranking reversal} for the pair $(A, B)$ occurs when $\Delta_1 \cdot \Delta_2 < 0$.

\subsection{Reversal criterion}

A ranking reversal occurs if and only if:
\begin{equation}
\mathrm{sign}(\Delta_1) \cdot \delta\Delta < -|\Delta_1|.
\end{equation}
This simply restates that a reversal requires two simultaneous conditions: the protocol shift must \emph{oppose} the initial ordering, and the magnitude of the shift must \emph{exceed} the initial margin. While straightforward, this criterion is useful as a diagnostic because it separates the direction of instability from its magnitude and motivates the decompositions below.

\subsection{Candidate-set decomposition}

For a fixed mapping policy and reference, consider a shift from candidate set $S_1$ to $S_2$. Write the margin in product form $\Delta(S) = b(S) \cdot g(S)$, where $b(S)$ is the candidate-set base rate (fraction of positives in $S$) and $g(S)$ is the base-rate-normalized discrimination gap between methods $A$ and $B$.

For the shift $S_1 \to S_2$:
\begin{equation}
\label{eq:decomposition}
\Delta_2 - \Delta_1 = \underbrace{(b_2 - b_1) \cdot g_1}_{\text{base-rate term}} + \underbrace{b_2 \cdot (g_2 - g_1)}_{\text{discrimination term}}.
\end{equation}

This decomposition separates a purely mechanical effect (the base rate changes when candidate sets differ in size and composition) from a substantive effect (the methods' relative discrimination changes in the new candidate space). If the normalized discrimination gap is invariant to candidate-set shift ($g_2 = g_1$) and the base rate remains positive, then base-rate scaling alone cannot reverse the ordering. Observing a reversal therefore implies that at least the discrimination term has shifted.

\subsection{Instability-region screening}

If $|\delta\Delta| \leq B$ for all shifts in a family, then all method pairs with $|\Delta_1| \leq B$ lie in an \emph{instability region} where reversal is possible under at least one shift in the family. This provides a practical screening tool: given a bound $B$ on the maximum margin shift observed across a shift family, any pair whose initial margin is within $B$ is flagged as potentially unstable. The criterion is designed for high recall (few missed reversals) at the cost of moderate precision.

\subsection{Mapping decomposition}

For mapping-policy shifts (changes in how gene identifiers are resolved), write $\mathcal{M} = c \cdot q$, where $c$ is the coverage (fraction of predictions that overlap with the reference) and $q$ is coverage-adjusted quality:
\begin{equation}
\mathcal{M}_2 - \mathcal{M}_1 = (c_2 - c_1) \cdot q_1 + c_2 \cdot (q_2 - q_1).
\end{equation}
This separates the effect of symbol-resolution changes on overlap from their effect on prediction quality within the overlapping set.

\section{Methods}
\label{sec:methods}

\subsection{Data sources}

We use previously generated benchmark summaries and score-evaluation outputs from a GRN benchmarking pipeline applied to three tissues from the Tabula Sapiens atlas \citep{tabulasapiens2022}: kidney, lung, and immune. The evaluation objects include tissue-stratified method-by-candidate summaries, immune-baseline method scores across multiple reference networks, and probe-prior evaluations across mapping-policy variants.

These artifacts represent a common GRN benchmarking stack built around curated references (DoRothEA \citep{holland2020dorothea}, TRRUST \citep{han2018trrust}, OmniPath \citep{turei2021integrated}, and composite unions) and six baseline inference methods: scGPT attention \citep{cui2024scgpt}, GENIE3 \citep{huynhthu2010inferring}, GRNBoost2 \citep{moerman2019grnboost2}, SCENIC \citep{aibar2017scenic}, and random baselines.

\subsection{Analyses}

We compute the following across each protocol-shift axis:

\paragraph{Pairwise reversal rates.} For each shift $(s_1, s_2)$ and each method pair, we check whether $\Delta_1 \cdot \Delta_2 < 0$ and report the reversal fraction with Wilson confidence intervals \citep{agresti1998approximate}.

\paragraph{Decomposition terms.} For candidate-set shifts, we compute the base-rate and discrimination terms from Eq.~\ref{eq:decomposition} and verify algebraic closure (reconstruction error $< 10^{-10}$).

\paragraph{Instability screening.} We perform leave-one-tissue-out cross-validation of the instability-region criterion, sweeping the quantile threshold used to set $B$ and reporting precision, recall, specificity, and F1.

\paragraph{Permutation null.} We construct a null distribution for candidate-shift reversal rates by randomly permuting method scores 5{,}000 times, preserving the candidate-set structure but destroying method identity.

All rate estimates include Wilson binomial confidence intervals \citep{agresti1998approximate}. Decomposition identities are algebraic and verified numerically for closure error. Following recommended practice for benchmark studies \citep{demsar2006statistical, weber2019essential}, claims are constrained to protocol-conditional benchmarking behavior; we do not make causal biological claims.

\section{Results}
\label{sec:results}

Table~\ref{tab:summary} summarizes the reversal rates across all four protocol axes examined.

\begin{table}[t]
\centering
\caption{Summary of pairwise ranking reversal rates across four protocol-shift axes.}
\label{tab:summary}
\begin{tabular}{@{}lccc@{}}
\toprule
Protocol axis & Reversals & Rate (\%) & 95\% CI \\
\midrule
Candidate-set shift & 22/135 & 16.3 & 11.0--23.4 \\
Tissue shift & 26/135 & 19.3 & 13.5--26.7 \\
Reference-network shift & 34/106 & 32.1 & 24.0--41.5 \\
Mapping-policy shift & 0/165 & 0.0 & 0.0--2.3 \\
\bottomrule
\end{tabular}
\end{table}

\subsection{Candidate-set shifts produce nontrivial reversal rates}

\begin{figure}[t]
\begin{center}
\includegraphics[width=\linewidth]{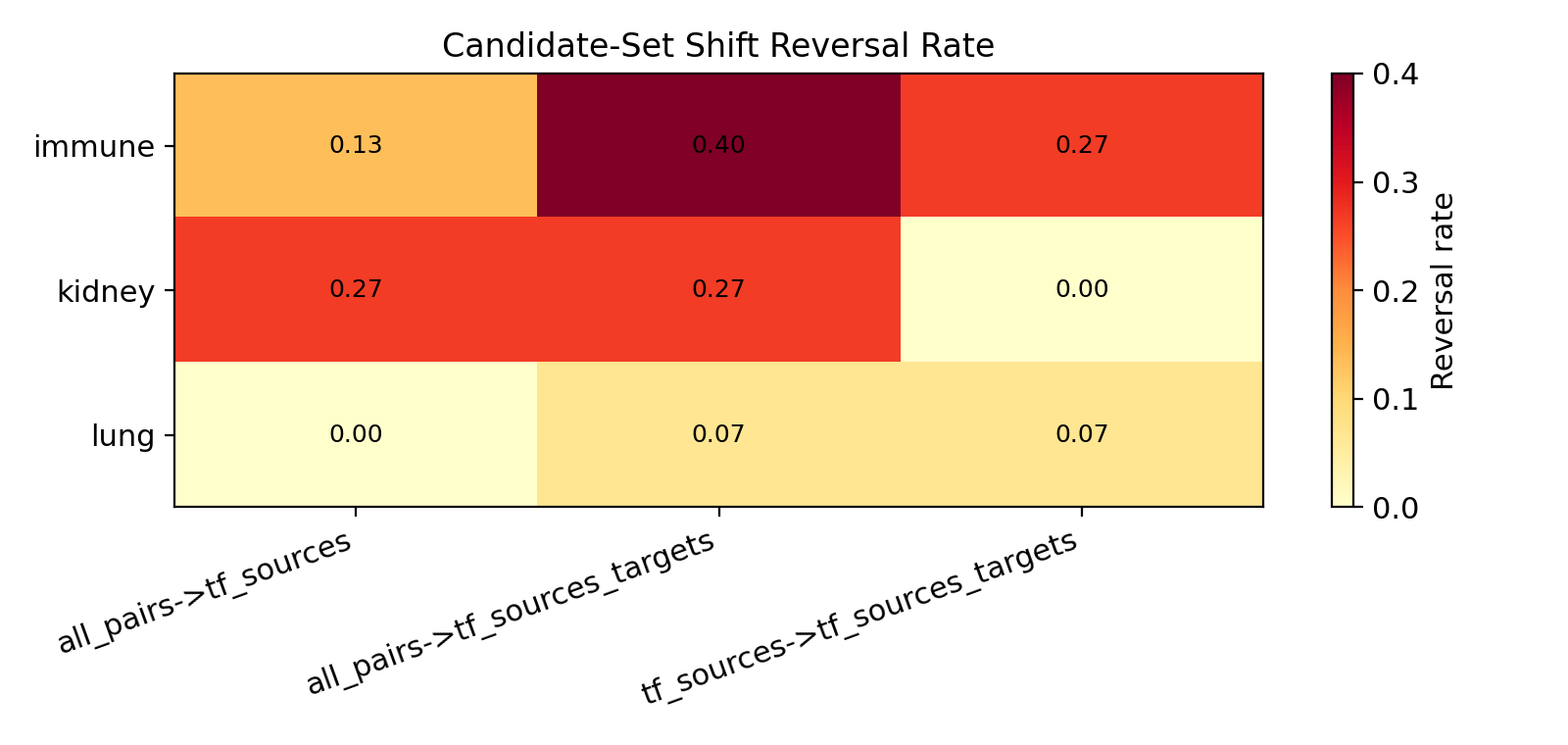}
\end{center}
\caption{Pairwise reversal rates under candidate-set shifts, stratified by tissue and shift type. Each cell shows the fraction of method pairs whose ranking reverses. Immune evaluations exhibit the highest sensitivity to candidate-set restriction.}
\label{fig:candidate_heatmap}
\end{figure}

Candidate-set shifts produce 22 out of 135 pairwise reversals (16.3\%, 95\% CI 11.0--23.4\%). The reversal rate is tissue-heterogeneous: immune evaluations shifting from all-pairs to TF-source-target candidate sets reach 40\% (6/15 pairs), while kidney evaluations shifting between TF-source and TF-source-target sets show 0/15 reversals (Figure~\ref{fig:candidate_heatmap}).

This result establishes that benchmark rank is not a method-intrinsic invariant but is protocol-conditional. A single leaderboard rank should not be used to justify biological claims without candidate-shift stability reporting.

\subsection{Reversals are discrimination-dominated, not base-rate-driven}

\begin{figure}[t]
\begin{center}
\includegraphics[width=\linewidth]{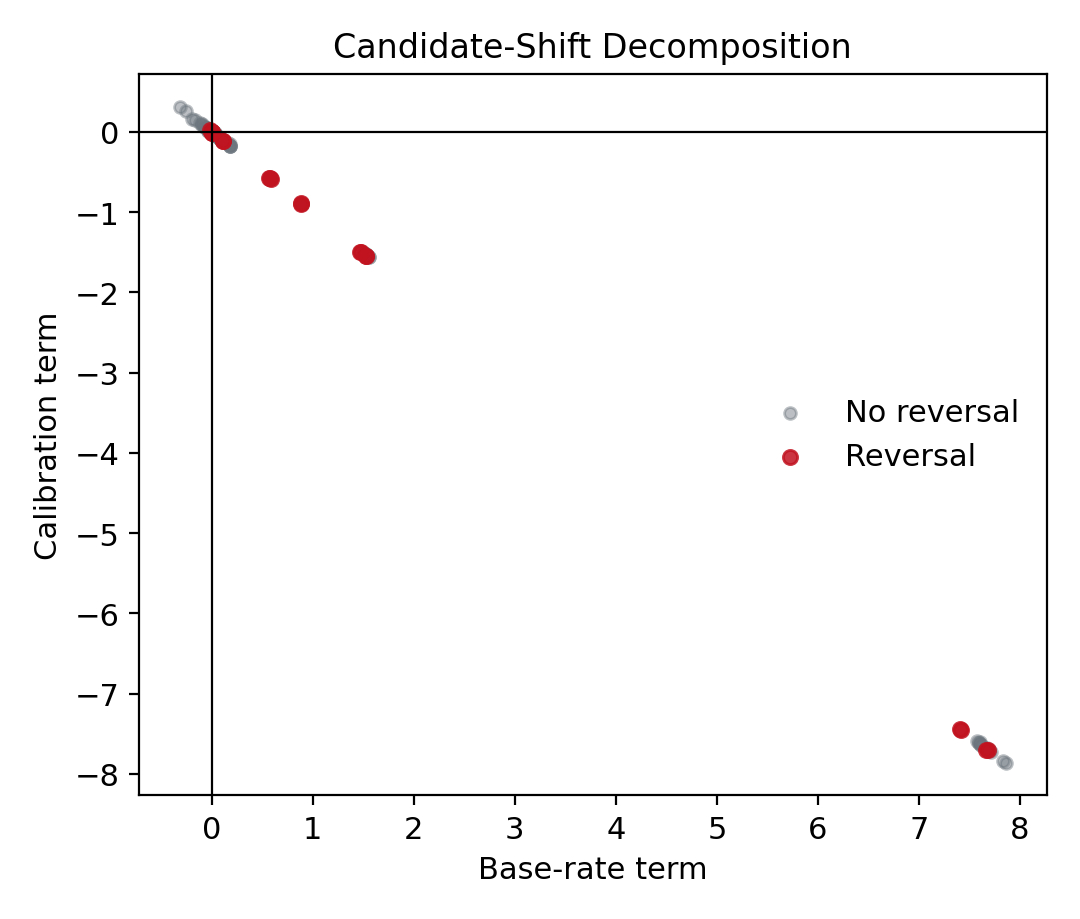}
\end{center}
\caption{Decomposition of margin shifts into base-rate and discrimination components. Reversal cases (red) show that discrimination changes, rather than base-rate inflation, drive rank flips.}
\label{fig:decomposition}
\end{figure}

Applying the decomposition from Eq.~\ref{eq:decomposition} to all candidate-shift reversal cases reveals a clear pattern: the discrimination term opposes the initial margin in 100\% of reversal cases, while the base-rate term opposes the initial margin in 0\% of cases. The mean ratio $|\text{discrimination}| / |\text{base-rate}|$ for reversal rows is 1.54 (Figure~\ref{fig:decomposition}).

This finding challenges a natural default assumption: that rank changes under candidate-set restriction are primarily driven by base-rate inflation (since restricting to known TF-target pairs mechanically increases the positive rate). Instead, the methods' relative ability to discriminate within the new candidate space changes, and these discrimination-shape shifts are the proximate cause of reversals. Benchmark designs that control for base rate (e.g., by normalizing metrics) will therefore not eliminate ranking instability.

\subsection{Tissue shifts amplify instability under constrained candidate spaces}

\begin{figure}[t]
\begin{center}
\includegraphics[width=\linewidth]{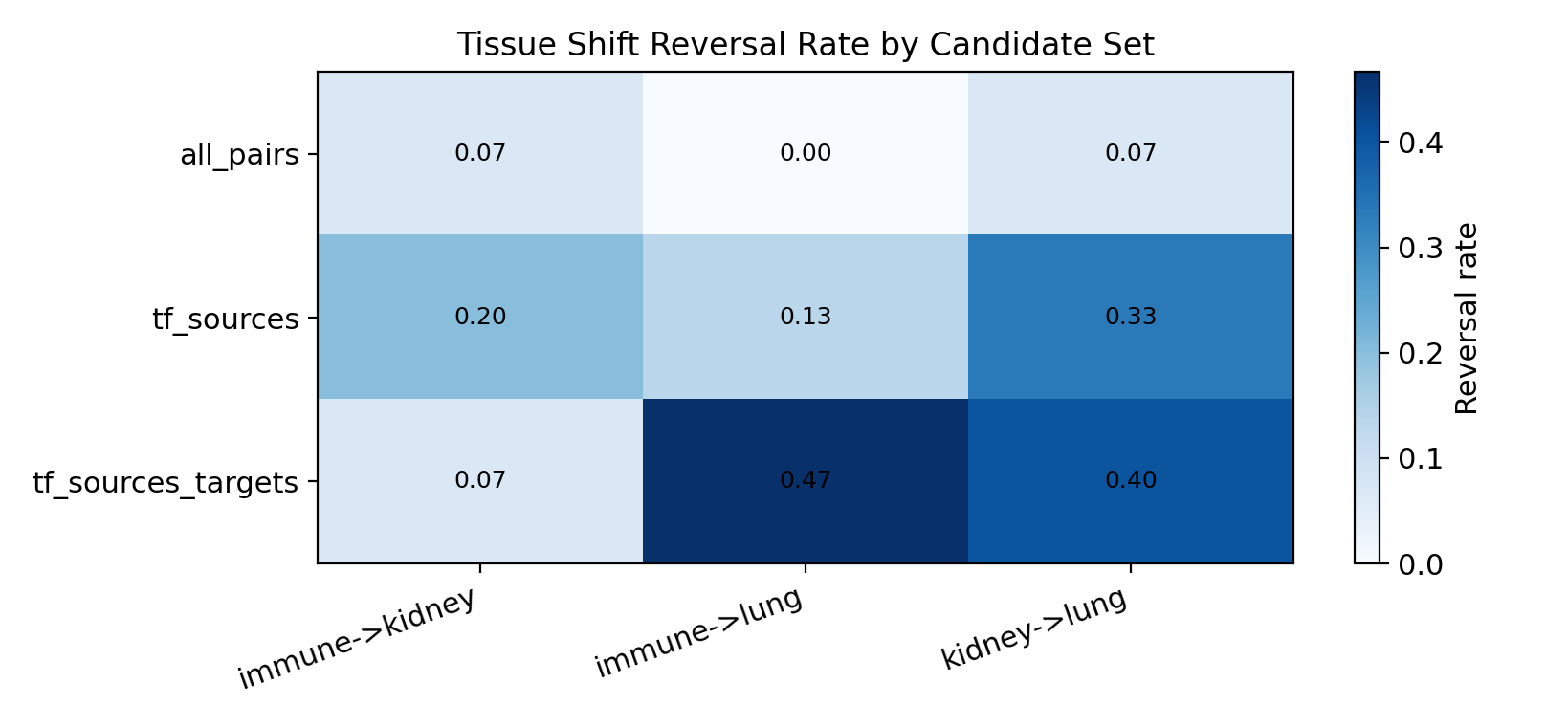}
\end{center}
\caption{Pairwise reversal rates under tissue shifts, stratified by candidate-set type. More constrained candidate spaces amplify cross-tissue ranking instability.}
\label{fig:tissue_heatmap}
\end{figure}

Tissue shifts yield 26 out of 135 reversals (19.3\%, 95\% CI 13.5--26.7\%). The reversal rate increases monotonically with candidate-space constraint: 4.4\% for all-pairs, 22.2\% for TF-source, and 31.1\% for TF-source-target evaluations (Figure~\ref{fig:tissue_heatmap}).

Cross-tissue rank transportability thus degrades as candidate spaces become more biologically curated. This is consistent with the interpretation that candidate constraints amplify tissue-specific regulatory-program mismatches by reducing the averaging effect of background edges.

\subsection{Reference-network shifts show the highest reversal rates}

\begin{figure}[t]
\begin{center}
\includegraphics[width=\linewidth]{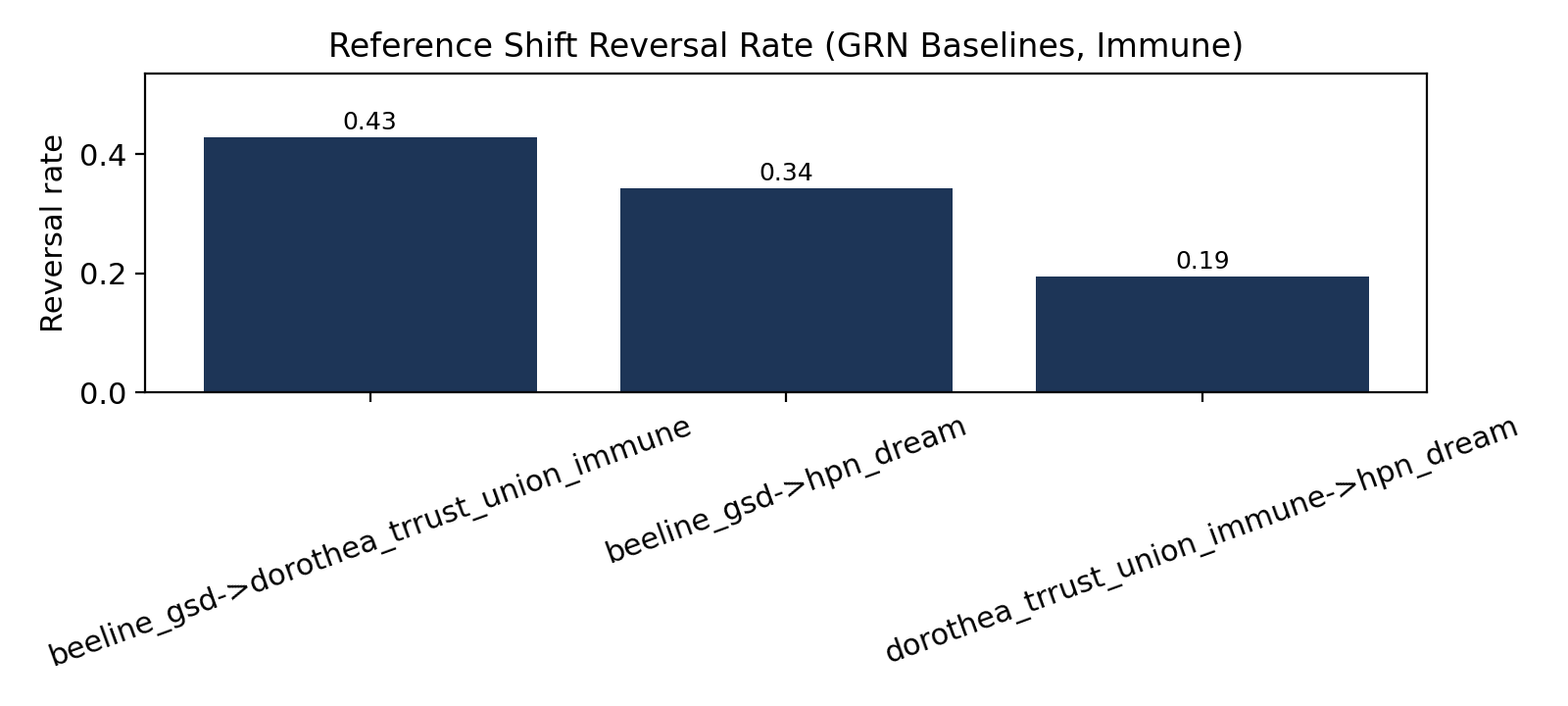}
\end{center}
\caption{Pairwise reversal rates under reference-network shifts in immune baseline evaluations. Different reference networks encode different biological evidence classes, producing high ranking instability.}
\label{fig:reference_bar}
\end{figure}

Reference shifts in immune baseline evaluations produce 34 out of 106 reversals (32.1\%, 95\% CI 24.0--41.5\%), the highest rate among the four protocol axes (Figure~\ref{fig:reference_bar}). The shift from Beeline GSD to the DoRothEA-TRRUST union reference reaches 42.9\% pairwise reversal.

This result implies that single-reference ``best method'' claims are likely overconfident, and multi-reference sensitivity analysis should be standard practice \citep{garcia2019benchmark}.

\subsection{Mapping-policy shifts preserve order despite large coverage changes}

Mapping-policy shifts (symbol normalization variants) show 0 out of 165 pairwise reversals (upper 95\% bound 2.28\%), despite mean coverage changes of +0.862. The mean F1 shift is negligibly small ($-4.32 \times 10^{-5}$).

In this subset, mapping behaves as an approximately order-preserving transform: it expands the set of scorable predictions without changing which method produces better-discriminating scores. Coverage changes still require reporting, however, because they alter the interpretability and comparability of absolute metric values even when relative ordering is preserved.

\subsection{Reversal structure is non-random}

\begin{figure}[t]
\begin{center}
\includegraphics[width=\linewidth]{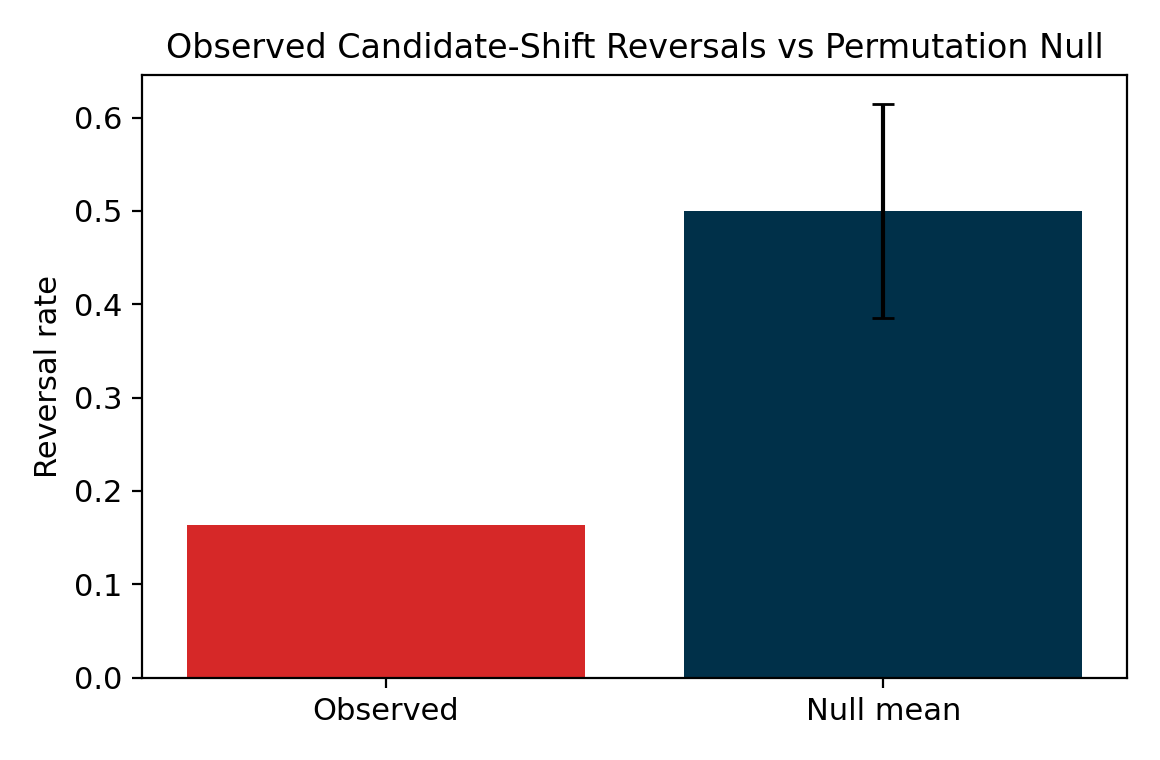}
\end{center}
\caption{Observed candidate-shift reversal rate versus permutation null distribution (5{,}000 permutations). The observed rate is far below random-order expectations, indicating substantial shared ranking structure that coexists with nontrivial instability.}
\label{fig:null}
\end{figure}

Under the 5{,}000-permutation null, the observed candidate-shift reversal rate of 0.163 falls far below the null mean of 0.500 (null 95\% interval: 0.385--0.615; Figure~\ref{fig:null}).

Rankings therefore retain substantial shared structure across protocol shifts, but with meaningful instability pockets. This intermediate regime---partially stable, partially unstable---is precisely the setting where formal stability diagnostics are most valuable.

\subsection{Instability screening provides useful triage}

\begin{figure}[t]
\begin{center}
\includegraphics[width=\linewidth]{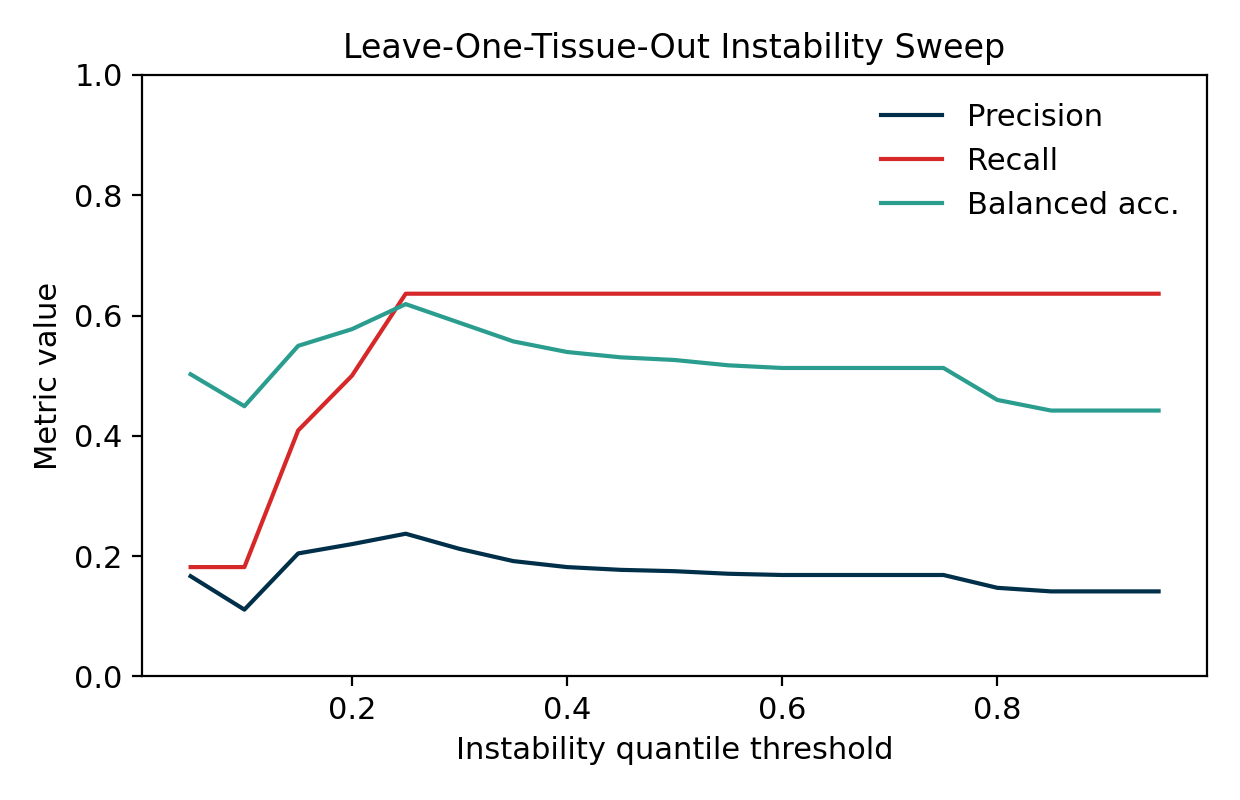}
\end{center}
\caption{Instability screening performance across quantile thresholds under leave-one-tissue-out cross-validation. The criterion provides useful high-recall triage at the cost of moderate precision.}
\label{fig:quantile}
\end{figure}

Leave-one-tissue-out evaluation of the instability-region criterion peaks near quantile threshold 0.25, achieving precision 0.237, recall 0.636, specificity 0.602, and F1 0.346 (Figure~\ref{fig:quantile}).

The instability-region diagnostic thus functions as a practical high-recall screening tool: it catches most true reversals while flagging a manageable number of false positives. It is not a stand-alone decision rule, but it can serve as a gating signal before expensive biological validation.

\section{Discussion}

\paragraph{Ranking instability is structured and decomposable.}
The reversal rates we observe (16--32\% depending on the protocol axis) are substantial enough to undermine single-protocol ranking claims, but they are also far below the $\sim$50\% that random ordering would produce. This intermediate regime is precisely the setting where formal stability diagnostics are most valuable, because neither blind trust nor blanket skepticism is warranted. Similar intermediate regimes have been documented in other benchmarking contexts \citep{recht2019imagenet, dehghani2021benchmark}.

\paragraph{Discrimination changes, not base-rate inflation, drive reversals.}
The decomposition analysis reveals that reversals under candidate-set shift are driven by discrimination changes rather than the mechanical base-rate inflation that candidate restriction produces. This means that methods respond differently to the \emph{composition} of the candidate space, not merely to its size. Benchmark designs that control for base rate (e.g., by normalizing metrics) will therefore not eliminate ranking instability. This finding parallels observations in broader ML evaluation that metric normalization does not remove evaluation-design sensitivity \citep{post2018call, marie2021scientific}.

\paragraph{Reference choice is the dominant instability source.}
Among the four protocol axes, reference-network shift produces the highest reversal rate (32.1\%). Different reference databases encode fundamentally different biological evidence classes---curated TF-target priors \citep{han2018trrust}, protein-protein interaction data \citep{szklarczyk2021string}, integrative signaling knowledge \citep{turei2021integrated}, and literature-mined associations \citep{holland2020dorothea}---and methods may legitimately perform differently against different evidence types. The practical implication is that single-reference evaluation is an underappreciated source of overconfident claims in GRN benchmarking, echoing concerns raised in the DREAM challenge literature \citep{marbach2012wisdom}.

\paragraph{Biological interpretation must be protocol-conditional.}
These results show that benchmark protocol partially defines the biological question being asked. Candidate-space restrictions emphasize particular regulator-target subsets, reference changes reweight biological evidence classes, and tissue context changes the active regulatory program. Observed method rank therefore mixes algorithmic quality with biological framing choices, and biological interpretation should be conditional on---and explicitly tied to---stability diagnostics. This is consistent with the broader principle that benchmark conclusions should be accompanied by sensitivity analyses \citep{weber2019essential, boulesteix2013plea}.

\paragraph{Practical recommendations.}
We propose three concrete reporting practices: (1)~evaluate methods across at least two candidate-set restrictions and report the reversal rate; (2)~include at least two reference networks and report reference-shift sensitivity; (3)~compute instability-region diagnostics as a standard supplement to metric tables. These are low-cost analyses that can be performed on existing benchmark outputs without rerunning inference, following the spirit of post-hoc benchmark auditing advocated by \citet{mangul2019systematic}.

\section{Limitations}

The present study relies on existing summary outputs rather than full replicate-level raw prediction matrices, which limits the granularity of uncertainty estimation at the individual-prediction level. The reference-shift analysis is currently centered on immune baseline outputs and would benefit from extension to kidney and lung tissues. The instability-screening diagnostic, while useful as a triage tool, has moderate precision and should not replace full multi-axis evaluation when computational resources allow. The scope is limited to six GRN inference methods across three tissues; broader method and tissue coverage---potentially including simulated data from tools like SERGIO \citep{dibaeinia2020sergio} or GeneNetWeaver \citep{schaffter2011genenetweaver}---would strengthen the generalizability of the observed reversal rates.

\section{Conclusion}

Ranking reversal is a first-order reliability concern for GRN benchmarking. We have provided a diagnostic framework---margin decompositions and practical screening tools---that makes ranking stability an explicit, quantifiable property of benchmark designs rather than an implicit assumption. The practical recommendation is direct: treat method rank as scientifically interpretable evidence only after cross-axis stability has been demonstrated.

\subsection*{Data and Code Availability}
All code, analysis scripts, figure-generation pipelines, and configuration files are available at \url{https://github.com/Biodyn-AI/grn-ranking-reversal-theory}. The repository includes instructions for reproducing all figures and tables from the provided benchmark summary data.

\appendix
\section{Permutation Null Construction}
\label{app:permutation}

The permutation null for candidate-shift reversal rates is constructed as follows. For each permutation $k \in \{1, \ldots, 5000\}$, we independently permute the vector of method scores within each candidate-set condition, preserving the number of methods and the score distribution but destroying the mapping between methods and scores. We then compute the pairwise reversal rate under the same shift definitions used for the observed data. The null distribution is the empirical distribution of these 5{,}000 reversal rates.

\section{Decomposition Algebra}
\label{app:decomposition}

The candidate-set decomposition (Eq.~\ref{eq:decomposition}) follows from the product rule:
\begin{align}
\Delta_2 - \Delta_1 &= b_2 g_2 - b_1 g_1 \nonumber \\
&= b_2 g_2 - b_1 g_2 + b_1 g_2 - b_1 g_1 \nonumber \\
&= (b_2 - b_1) g_2 + b_1 (g_2 - g_1).
\end{align}

The version in the main text uses $g_1$ in the base-rate term and $b_2$ in the discrimination term, corresponding to a different factorization of the same product-rule identity. Both decompositions are valid; we report both in our numerical analyses and verify that both reconstruct the total shift to within machine precision ($< 10^{-10}$).

\end{document}